\documentclass[%
 reprint,
superscriptaddress,
 amsmath,amssymb,
 prl,
]{revtex4-1}
\usepackage[acronym]{glossaries}
\newacronym{oam}{OAM}{orbital  angular momentum }
\usepackage{graphicx}% Include figure files
\usepackage{dcolumn}% Align table columns on decimal point
\usepackage{bm}
\usepackage{siunitx}
\usepackage{eufrak}% bold math
\begin{document}

\preprint{APS/123-QED}

\title{Orbital  angular momentum  coupling in elastic photon-photon scattering}% Force line breaks with \\

\author{R. Aboushelbaya}
\email{ramy.aboushelbaya@physics.ox.ac.uk}
\affiliation{Clarendon Laboratory, University of Oxford, Parks Road, Oxford OX1 3PU, United Kingdom}
\author{K. Glize}%
\affiliation{Central Laser Facility, STFC Rutherford Appleton Laboratory, Didcot OX11 0QX, United Kingdom}
\author{A. F. Savin}
\affiliation{Clarendon Laboratory, University of Oxford, Parks Road, Oxford OX1 3PU, United Kingdom}
\author{M. Mayr}
\affiliation{Clarendon Laboratory, University of Oxford, Parks Road, Oxford OX1 3PU, United Kingdom}
\author{B. Spiers}
\affiliation{Clarendon Laboratory, University of Oxford, Parks Road, Oxford OX1 3PU, United Kingdom}
\author{R. Wang}
\affiliation{Clarendon Laboratory, University of Oxford, Parks Road, Oxford OX1 3PU, United Kingdom}
\author{J. Collier}%
\affiliation{Central Laser Facility, STFC Rutherford Appleton Laboratory, Didcot OX11 0QX, United Kingdom}
\author{M. Marklund}%
\affiliation{Department of Physics, University of Gothenburg, SE-41296 Gothenburg, Sweden}
\author{R. M. G. M. Trines}%
\affiliation{Central Laser Facility, STFC Rutherford Appleton Laboratory, Didcot OX11 0QX, United Kingdom}
\author{R. Bingham}
\affiliation{Central Laser Facility, STFC Rutherford Appleton Laboratory, Didcot OX11 0QX, United Kingdom}
\affiliation{Department of Physics, University of Strathclyde, Strathclyde G4 0NG, United Kingdom}
\author{P. A. Norreys}
\affiliation{Central Laser Facility, STFC Rutherford Appleton Laboratory, Didcot OX11 0QX, United Kingdom}
\affiliation{Clarendon Laboratory, University of Oxford, Parks Road, Oxford OX1 3PU, United Kingdom}
%




\begin{abstract}
In this letter, we investigate the effect of orbital  angular momentum  (OAM) on elastic photon-photon scattering in vacuum for the first time. We define exact solutions to the vacuum electromagnetic wave equation which carry OAM. Using those, the expected coupling between three initial waves is derived in the framework of an effective field theory based on the Euler-Heisenberg Lagrangian and shows that OAM adds a signature to the generated photons thereby greatly improving the signal-to-noise ratio. This forms the basis for a proposed high-power laser experiment utilizing quantum optics techniques to filter the generated photons based on their OAM state.  
%\begin{description}
%\end{description}
\end{abstract}

\maketitle

%\tableofcontents

Photon-photon scattering in vacuum is an effect that has attracted considerable interest over the past few decades \cite{brodin_proposal_2001,kaplan_field-gradient-induced_2000}. In classical electromagnetism, the linearity of Maxwell's equations indicates that scattering between photons cannot happen in the vacuum. Nevertheless, results from quantum electrodynamics (QED) have shown that the vacuum is, in fact, non-linear. This means that the aforementioned scattering is actually possible, albeit very difficult to observe \cite{berestetskii_quantum_1982}. Many different proposals have been made to detect such processes using microwave cavities \cite{brodin_proposal_2001}, plasma channels \cite{shen_photonphoton_2003} and high-power lasers \cite{lundstrom_using_2006, king}. The main obstacle to these schemes has been the relatively weak predicted signal compared to the noise inherent to any experiment. As such, scattering between real photons has yet to be observed in the vacuum.

In this letter, we investigate the effect of \gls{oam} coupling in elastic photon-photon scattering and we show, for the first time, that \gls{oam} contributes an additional unique signature that will help filter the background noise and is thus predicted to make the detection of these weak signals easier. \gls{oam} is a relatively recently understood degree of freedom of light that is distinct from the more widely understood spin angular momentum (SAM) which is related to light's polarization \cite{torres_twisted_2011}. It appears in electromagnetic modes which have an azimuthal phase dependence, $E(\mathbf{r})\propto e^{il\phi}$, where conventionally $l$ denotes the \gls{oam} state of the mode. It should be noted that $l$ here represents the projection of \gls{oam} on the direction of propagation in contrast to its use in the quantum mechanical treatment of bound electronic states where it represents the absolute value of the ``total" \gls{oam}. Interest in researching these modes has grown ever since they were shown to carry a quantized net angular momentum separate from their spin \cite{allen_orbital_1992}. Since it is unbounded and can have any integer value, \gls{oam} has many potential applications in different fields including: enhanced imaging techniques, light manipulation and high density coding of information for communication \cite{torres_twisted_2011,lei_fast-switchable_2017,Ghaleh:18}. Consequentially, over the course of the previous two decades, the techniques required to efficiently generate \gls{oam} states, and detect and filter photons according to their \gls{oam} state (right down to the single-photon level) have been honed to a great level of precision \cite{hadfield_single-photon_2009,leach_measuring_2002}. For this reason, we are interested in investigating its impact on photon-photon scattering. 

This letter will be organized as follows: first, exact solutions to the electromagnetic wave equation which carry \gls{oam} will be discussed. These modes will then be used to investigate the effect of \gls{oam} on photon-photon interaction within the framework of an effective field theory \cite{heisenberg_folgerungen_1936}. Finally, a proposed experimental configuration will be presented that is expected to validate the accompanying theoretical work.

The simplest solution to the linear electromagnetic wave equation is the plane wave decomposition, where a plane wave is described by:
\begin{equation}
    \textbf{E}(\textbf{r},t) = \mathcal{E}_0e^{i(\textbf{k}\cdotp\textbf{r}-\omega t)}\hat{\textbf{n}}
\end{equation}
where $\mathbf{k}$ is the wavevector, $\omega$ is the angular frequency and $\hat{\mathbf{n}}$ is the polarization unit vector of the plane wave which necessarily satisfies the transversality condition ($\textbf{k}\cdotp\hat{\textbf{n}} = 0$) arising from Maxwell's equations. We then consider an infinite superposition of plane waves whose wave vectors lie on the surface of a cone with half-angle $\alpha$ around the z-axix:

\begin{equation}
    \textbf{E}_l(\textbf{r},t) = \dfrac{\mathcal{E}_0}{2\pi}\int_0^{2\pi}(i)^le^{i(\textbf{k}(\phi_k)\cdotp\textbf{r}-\omega_0 t+l\phi_k)}\hat{\textbf{n}}(\phi_k)d\phi_k
\label{eq:super}
\end{equation}
\begin{equation}
    \textbf{k}(\phi_k) = k\cos(\alpha)\hat{\textbf{z}} - k\sin(\alpha)(\cos(\phi_k)\hat{\textbf{x}} + \sin(\phi_k)\hat{\textbf{y}})  
\end{equation}
\begin{equation}
    \hat{\textbf{n}}(\phi_k) = \sin(\phi_k)\hat{\textbf{x}} - \cos(\phi_k)\hat{\textbf{y}}
\end{equation}
where $l\in\mathbb{Z}$ is the azimuthal mode number. $\mathbf{E}_l$ is obviously then an exact solution to the electromagnetic wave equation as it is a linear superposition of exact solutions. Its integral representation can be simplified by decomposing the trigonometric functions into their exponential form and using the fact that $(1/2\pi)\int_0^{2\pi}e^{i(l\phi-x\sin(\phi))}d\phi=J_l(x)$ \cite{abramowitz_handbook_1974}, where $J_l(x)$ is the $l$-th order Bessel function of the first kind. Thus, the electric field in Eq.(\ref{eq:super}) can be rewritten in cylindrical coordinates ($z, \rho, \phi$) as: 
\begin{equation}
\begin{split}
    \textbf{E}_l(z,\rho,\phi,t)= & \mathcal{E}_0e^{i(\kappa z-\omega t+l\phi)}\Bigg[- \dfrac{lJ_l(\beta\rho)}{\beta\rho}\hat{\mathbf{e}}_\rho \\
    & +i\dfrac{J_{l+1}(\beta\rho)-J_{l-1}(\beta\rho)}{2}\hat{\mathbf{e}}_\phi\Bigg]
\end{split}
\label{eq:bessel}
\end{equation}
Where $\kappa = k\cos(\alpha)$ and $\beta = k\sin(\alpha)$. The corresponding magnetic field can then be calculated from Eq.(5) using Maxwell's equations ($\nabla\times\mathbf{E}_l = -\partial\mathbf{B}_l/\partial t$). This solution can be seen as an \gls{oam}-carrying extension of the electromagnetic Bessel modes. One of the interesting properties of Bessel modes is that they propagate without diffraction as can be seen by noting that the radial profile of the field is independent of the longitudinal position along the direction of propagation. True Bessel beams are unphysical as they carry infinite energy. However, Bessel-like modes, which retain some of the aforementioned interesting properties over a known coherence length, have been generated using axicon lenses \cite{doi:10.1080/0010751042000275259}. When considering the case of $l\neq0$, the field has an azimuthal phase dependence, and therefore as shown in \cite{allen_orbital_1992}, it carries nonzero \gls{oam}. This can be confirmed by calculating the time-averaged angular momentum density of the field $\langle\mathfrak{L}\rangle=(1/2)\mathbf{r}\times\Re{(\textbf{E}^*\times\textbf{B})}$.
\begin{equation}
\begin{split}
    \langle\mathfrak{L}\rangle =& \dfrac{\epsilon_0\mathcal{E}_0^2}{ c}\Bigg(\dfrac{l}{k_0}J^2_l(\beta\rho)\hat{\mathbf{z}} -\dfrac{lz\sin(\alpha)}{\beta\rho}J^2_l(\beta\rho)\hat{\mathbf{\rho}}\\&-\rho\cos(\alpha)\dfrac{J^2_{l+1}(\beta\rho)-J^2_{l-1}(\beta\rho)}{2}\hat{\mathbf{\phi}}\Bigg)
\end{split}
\end{equation}
If $\langle\mathfrak{L}\rangle$ is integrated over a finite circular transverse profile of the field, only the z-component of the angular momentum density contributes as the polar unit vectors integrate to 0. This results in a total angular momentum $\langle\mathbf{L}\rangle\propto l\hat{\mathbf{z}}$.

It should be noted that this method of defining light modes that carry \gls{oam} differs from the standard approach of considering Laguerre-Gaussian (LG) modes \cite{allen_orbital_1992}. The method used here offers two advantages, the first being that the field defined in Eq.(\ref{eq:bessel}) is, by construction, divergence-free and hence an exact solution of Maxwell's equations. Secondly, it offers a simple representation of \gls{oam} beams as a superposition of plane waves which proves highly advantageous when calculating the nonlinear interaction terms compared to the LG modes which contain a complicated envelope that makes the calculation intractable.  

Classically, photon-photon scattering is physically impossible since Maxwell's equations are linear \cite{maxwell_dynamical_1865}, meaning that any superposition of solutions to the wave equation is itself a solution. However, the full quantum electrodynamic treatment of the vacuum field predicts the existence of vacuum polarization and the possibility of photon-photon scattering mediated by virtual electron-positron pairs \cite{berestetskii_quantum_1982}. The exact treatment of this problem is quite challenging, however, the interaction can be approximated by the Euler-Heisenberg Lagrangian which accounts for this vacuum polarization up to one loop in the corresponding Feynman diagram, thereby accounting for the lowest order contribution to the photon-photon QED scattering amplitude. It can be written in Gaussian units, in the limit for fields oscillating slower than the Compton frequency, as \cite{heisenberg_folgerungen_1936}:
\begin{equation}
    \mathcal{L} = \dfrac{1}{8\pi}\Big((\mathbf{E}^2-\mathbf{B}^2) + \xi[(\mathbf{E}^2-\mathbf{B}^2)^2+ 7(\mathbf{E}\cdotp\mathbf{B})^2]\Big)
\end{equation}
Where $\xi=\dfrac{\hbar e^4}{45 \pi^2 m^4 c^7}$. A modified electromagnetic wave equation can then be derived using the Euler-Lagrange equations. The quartic corrections to the Lagrangian lead to a non-linear source term for the wave equation, even in the vacuum. 
\begin{equation}
    \partial^\mu\partial_\mu\mathbf{E} = 4\pi(c^2\mathbf{\nabla}(\mathbf{\nabla}\cdotp\mathbf{P}) - \partial_t(\partial_t\mathbf{P} + c\mathbf{\nabla}\times\mathbf{M}))
\end{equation}
Where $\partial^\mu\partial_\mu=\partial^2_t-c^2\nabla^2$, and $\mathbf{M}$ and $\mathbf{P}$ are the effective vacuum magnetization and polarization respectively, described by: 
\begin{equation}
    \mathbf{M} = (4\pi)^{-1}\xi(-2(\mathbf{E}^2-\mathbf{B}^2)\mathbf{B} + 7(\mathbf{E}\cdotp\mathbf{B})\mathbf{E}) 
\end{equation}
\begin{equation}
    \mathbf{P} = (4\pi)^{-1}\xi(2(\mathbf{E}^2-\mathbf{B}^2)\mathbf{E} + 7(\mathbf{E}\cdotp\mathbf{B})\mathbf{B}) 
\end{equation}
These equations indicate that the source term in Eq.(8) has a non-linear cubic dependence on the electromagnetic fields. Therefore, if we start with three plane waves $(\mathbf{k}_i, \omega_i)_{i\in\{1,2,3\}}$, it is clear that the three modes will mix generating a fourth wave ($\mathbf{k}_4, \omega_4$) that satisfies the following energy and momentum matching conditions: 
\begin{equation}
    \mathbf{k}_1+\mathbf{k}_2=\mathbf{k}_3+\mathbf{k}_4 \ \ \ \ \ \ \ \omega_1+\omega_2=\omega_3+\omega_4
\label{eq:phasematching}
\end{equation}
There are, of course, many other interaction terms that involve different combinations of the three initial waves. However, in the rest of this letter, only terms that correspond to the matching conditions in Eq.(11) will be considered. This is because it is the most convenient generated wave to measure experimentally and allows for flexibility in the geometrical setup of the incoming beams as will be seen in the following parts. This does not necessarily mean that the other interactions terms are particularly weaker. However, they are also not any stronger, and waves generated through them will have frequencies and directions of propagation (and different OAM state as will be seen later on) different from Eq.(\ref{eq:phasematching}) meaning that they can safely be neglected when estimating the number of detected photons in the experimental geometry that will be considered later on.

Let the three initial plane waves be $\mathbf{E}_1 = E_1e^{i(k_0x-\omega_0t)}\hat{\mathbf{y}}$, $\mathbf{E}_2 = E_2e^{i(-k_0x-\omega_0t)}\hat{\mathbf{z}}$ and $\mathbf{E}_3 =(i)^l E_3e^{i(\mathbf{k}_3(\phi_k)\cdotp\mathbf{r}-\omega_0t)}e^{il\phi_k}\hat{\mathbf{n}}(\phi_k)$ where $\mathbf{k}_3(\phi_k)= k_0\cos(\alpha)\hat{\textbf{z}} - k_0\sin(\alpha)(\cos(\phi_k)\hat{\textbf{x}} + \sin(\phi_k)\hat{\textbf{y}})$ and $\hat{\textbf{n}}(\phi_k) = \sin(\phi_k)\hat{\textbf{x}} - \cos(\phi_k)\hat{\textbf{y}}$. It is clear that $\mathbf{E}_3$ is a plane wave that lies on the surface of a cone with half-angle $\alpha$ similar to the aforementioned case. Further, $\alpha$ is considered to be very small, which is a reasonable assumption with Bessel beams as they are generally generated with axicon lenses which have a very shallow angle. Keeping only the interaction terms of Eq.(8) that are resonant for the generated fourth wave $\mathbf{E}_4$. The non-linear wave equation can then be written as \cite{supplementary_material}:
\begin{equation}\label{eq:source}
    \partial^\mu\partial_\mu\mathbf{E}_4 \approx (-i)^l\omega_0^2\xi\hat{\mathbf{v}}(\phi_k)E_1E_2E_3^*e^{i(\mathbf{k}_4(\phi_k)\cdotp\mathbf{r}-\omega_0t-l\phi_k)}
\end{equation}
where $\hat{\mathbf{v}}(\phi_k)=\cos(\alpha)\cos(\phi_k)\hat{\mathbf{x}}+\cos(\alpha)\sin(\phi_k)\hat{\mathbf{y}}+\sin(\alpha)\hat{\mathbf{z}}$ and $\mathbf{k}_4=-\mathbf{k}_3$.  

Consider now the case where $\mathbf{E}_3$ is not a plane wave, but an $l$-th order \gls{oam} mode whose average wavevector is directed along the positive z-axis. As seen above, the \gls{oam} beam can be decomposed into an appropriate superposition of plane waves. It is only then necessary to perform an integration on Eq.(\ref{eq:source}), similar to the one presented above in Eq.(2), and find that the source term, in this case, can be rewritten as Bessel functions with an overall azimuthal phase dependence. 
\begin{equation}
     \partial^\mu\partial_\mu\mathbf{E}_4 \approx 
    \omega_0^2\xi E_1E_2E_3^*\mathbf{J}_l(\rho, \phi)e^{i(-\kappa z-\omega_0t - l\phi)}
\end{equation}
where $\beta=k_0\cos(\alpha)$, $\kappa=k_0\sin(\alpha)$ and $\mathbf{J}$ is a vector containing the transverse radial dependence of the source.  
\begin{equation}
\begin{split}
     \mathbf{J}_l(\rho,\phi)=&\sin(\alpha)J_l(\beta\rho)\hat{\mathbf{z}}+i\cos(\alpha)\dfrac{J_{l+1}(\beta\rho)-J_{l-1}(\beta\rho)}{2}\hat{\mathbf{\rho}}\\&-\dfrac{l\cos(\alpha)J_l(\beta\rho)}{\beta\rho}\hat{\mathbf{\phi}}
\end{split}
\end{equation}
Eq.(13) can be solved using the standard Green's function method, and the generated field, far from the source, can be written as: 
\begin{equation}
    \mathbf{E}_4=\dfrac{(i)^l}{4\pi rc^2}\omega_0^2\xi E_1E_2E_3^*\mathbf{\Lambda}_l(\theta)e^{i(k_0r-\omega_0t-l\phi)}
    \label{eq:generated_field}
\end{equation}
where:
\begin{equation}
\begin{split}
    \Lambda_l(\theta) =&2\pi(\sin(\alpha)A_l\hat{\mathbf{z}}- \cos(\alpha)\dfrac{A_{l+1}+A_{l-1}}{2}\hat{\mathbf{\rho}}\\&-i\cos(\alpha)\dfrac{A_{l+1}-A_{l-1}}{2}\hat{\mathbf{\phi}})
\end{split}
\end{equation}
\begin{equation}
    A_l = \int\int_{V'}e^{-i(\kappa+k_0\cos(\theta))} J_{-l}(k_0\sin(\alpha)\rho')J_l(\beta\rho')\rho'd\rho'dz'
\end{equation}
and $r\in\mathbb{R}^+$ is the spherical radial distance from the origin of the finite interaction volume $V'$ whose extent will depend on the size of the focal spots of the three beams being used in the experimental setup. The interaction volume is artificially limited in this calculation because the beams used in the analytical derivation have infinite extent and are non-integrable over all of space, while physical beams will always have an integrable intensity envelope. 
\begin{figure}
    \centering
    \includegraphics[width=\linewidth]{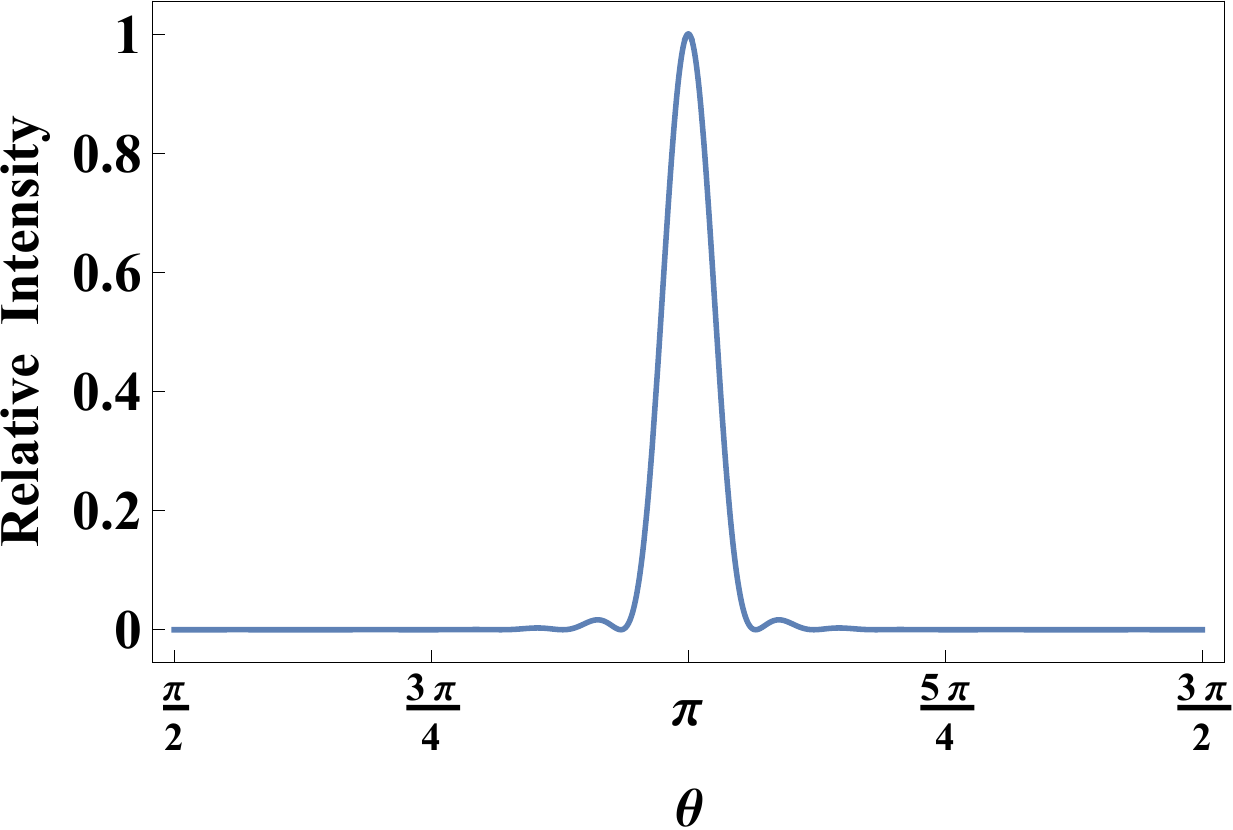}
    \caption{Plot of the angular dependence of the field of the generated photons $|\mathbf{E}_4|^2$ to show the direction of propagation of the generated wave. $\theta$ is the polar angle defined, as it usually is in the spherical coordinate system, from the z-axis.}
    \label{fig:curve}
\end{figure} 
The direction of propagation of this generated wave can be seen in the angular dependence of $\mathbf{\Lambda}_l(\theta)$ in Fig.(\ref{fig:curve}). From it we can conclude that the generated wave propagates along the same axis but in the opposite direction as the initial \gls{oam} mode $\mathbf{E}_3$, which is intuitive as $\mathbf{E}_1$ and $\mathbf{E}_2$ are anti-parallel, and linear momentum is conserved. It can also be clearly seen in Eq.(15) that the generated field has the opposite azimuthal phase dependence of $\mathbf{E}_3$. This means that the \gls{oam} of the generated photons will be opposite in sign, which is to be expected as a consequence of conservation of angular momentum since the two other intial waves $\mathbf{E}_1$ and $\mathbf{E}_2$ were plane waves. We can then use Eq.(15) to calculate the generated intensity $I_4=(1/2)c\epsilon_0|E_4|^2$ which can be integrated over the surface of the interaction volume to calculate the number of generated photons for specific initial laser parameters.

Eq.(\ref{eq:generated_field}) shows that for a specific set of three-beam geometries, if one of the lasers is carrying some \gls{oam} ($l$) then the photons generated due to photon-photon scattering will carry the opposite value $l_4=-l$ as can be seen from the azimuthal dependence in the phase of the generated field. We propose to use this additional signature to design a novel high-power laser experiment to verify this hitherto undetected phenomenon. In order to make the generated photons even easier to distinguish, the geometry will be modified slightly from the one considered previously. Consider instead three initial waves, where ($\mathbf{k}_1=-2(\omega_0/c)(1/2\hat{\mathbf{z}}+\sqrt{3}/2\hat{\mathbf{x}})$, $\hat{\textbf{n}}_1=\hat{\mathbf{y}}$) and ($\mathbf{k}_2=-2(\omega_0/c)(1/2\hat{\mathbf{z}}-\sqrt{3}/2\hat{\mathbf{x}})$, $\hat{\textbf{n}}_2=\hat{\mathbf{y}}$) are plane waves, and $\mathbf{E}_3$ is an \gls{oam} beam with $l_3 = 1$ and $\omega_3=\omega_0$ that is propagating along the positive z-axis. These kinds of modes can be efficiently generated in the laboratory using mode-conversion methods such as so-called spiral phase mirors (SPM) which have efficiently generated OAM-carrying high power lasers \cite{longman_aps}. According to Eq.(\ref{eq:phasematching}) and Eq.(\ref{eq:generated_field}), we should then expect to detect photons generated at $\omega_4=\omega_1+\omega_2-\omega_3=3\omega_0$ with $l_4=-1$ propagating along the negative z direction. These generated photons are then distinct, from any of the initial radiation in the interaction, in frequency, direction of travel and \gls{oam} state. The expected number of photons generated by photon-photon scattering $N_\gamma$ can be calculated from Eq.(15) using:
\begin{equation}
N_\gamma = \dfrac{\epsilon_0c\tau}{2\hbar\omega_4}\int_0^{\pi}\int_0^{2\pi}|E_4(r,\theta,\phi)|^2r^2\sin(\theta)d\theta d\phi 
\end{equation} 
\begin{figure}
    \centering
    \includegraphics[width=\linewidth]{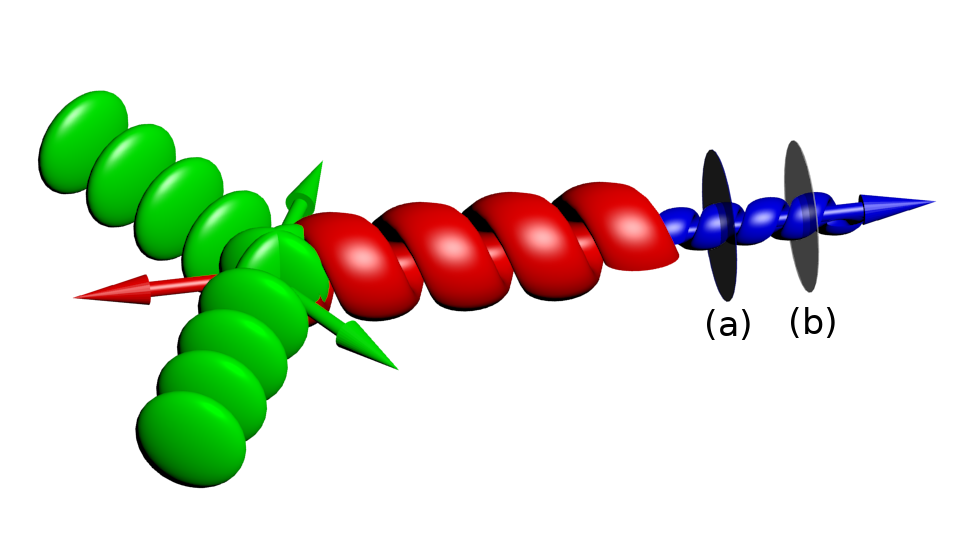}
    \caption{A sketch of the proposed geometry for the experiment showing the three main initial beams. The $2\omega_0$ (green) beams have a flat phase front whereas the $\omega_0$ (red) beam has a helical front arising from the nonzero \gls{oam}. The figure illustrates the generated $3\omega_0$ (blue) spiral photons. (a) and (b) are the frequency and \gls{oam} filters respectively, which will be used to remove background noise.}
    \label{fig:geometry}
\end{figure}
where $\tau$ is the duration of the interaction. In the case of a high-power laser of \SI{10}{\peta\watt} at a wavelength of \SI{800}{\nano\meter}, where the laser pulse duration is $\tau=\SI{30}{\femto\second}$, the beam must be split into three, where two of the component parts are frequency doubled. Considering a second harmonic conversion efficiency of 30\% and a spot size of \SI{5}{\micro\meter}, Eq.(18) predicts that $N_\gamma$ is expected to be on the order of 100 photons per shot. It should be noted that since unphysical light modes were used for this derivation, it can only provide an initial estimate of the number of generated photons through photon-photon scattering. This is, however, not a major concern as any additional realistic envelope (such as a Gaussian envelope) will vary slowly compared to the oscillations of the field and as such will have negligible contributions to the photon-photon scattering interaction. Fig.(\ref{fig:photons}) shows the scaling of $N_\gamma$ with respect to the power of the initial laser, while keeping all other parameters the same as the above example. Comparing this scaling to the one in the case with no OAM \cite{lundstrom_using_2006}, it is clear that the introduction of \gls{oam} does not change the coupling strength of the photon-photon interaction and the predicted number of photons generated from it is of the same order when lasers parameters are matched. 
\begin{figure}
    \centering
    \includegraphics[width=\linewidth]{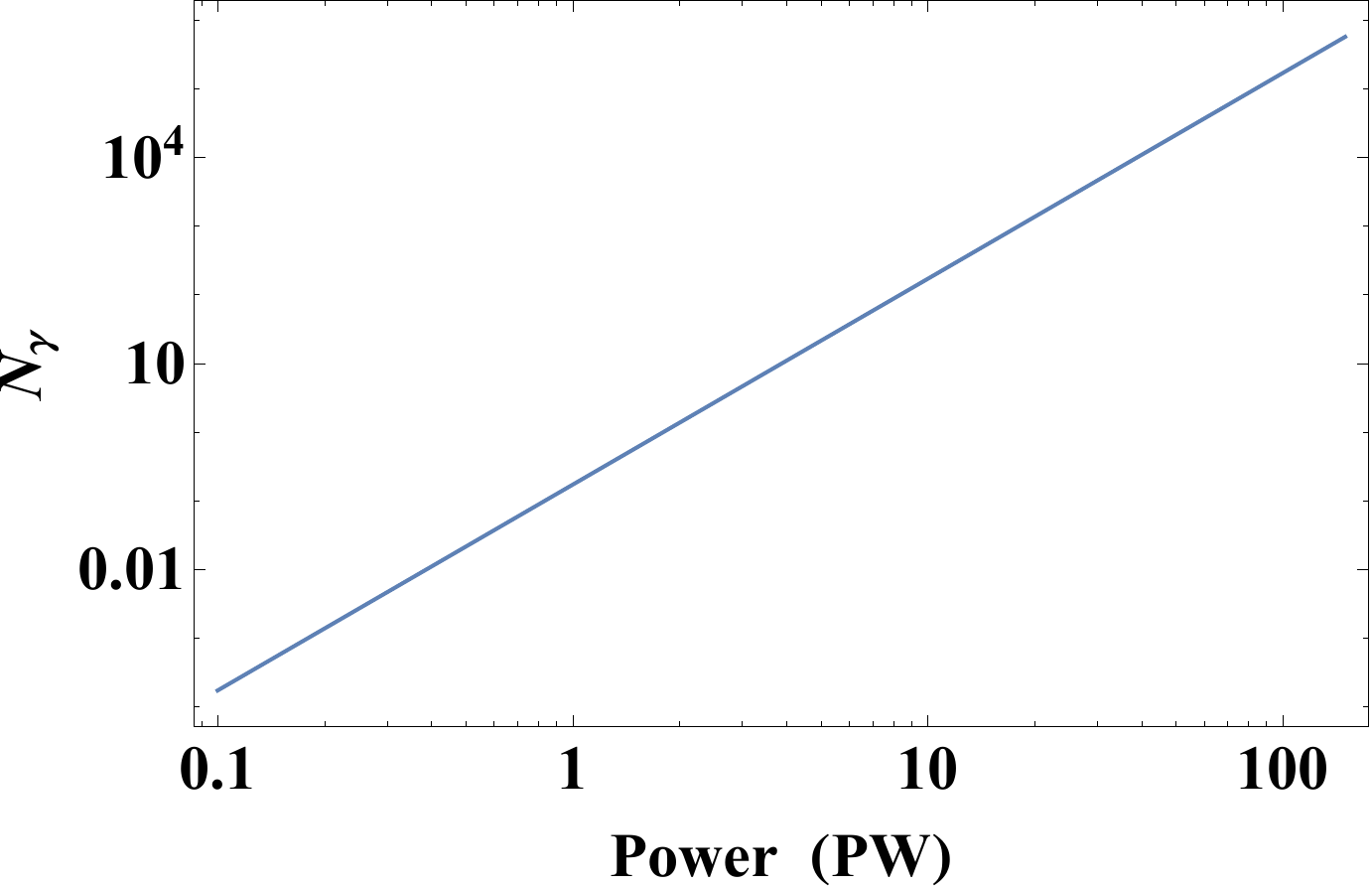}
    \caption{Plot showing the scaling the estimated number of generated photons as a function of initial laser power while fixing all other laser parameters (focal spot and pulse duration).}
    \label{fig:photons}
\end{figure}

Although the number of photons may seem small, especially compared to the other 3 high-power beams (e.g. the OAM-carrying beam contains $\mathrm{O}(10^{20})$ photons), its multiple different signatures will allow for efficient filtering of any other background radiation. There have been many successful experiments where even single photons have been filtered by their \gls{oam} state and detected using highly efficient single-photon detectors \cite{nicolas_quantum_2014,leach_measuring_2002}. Along with \gls{oam} filtering, frequency filtering with ultra narrow-band bandpass filters would block any photons whose frequency is sufficiently different from what is expected from the previously defined three-beam interaction. The single-photon detectors can also be gated to only be sensitive to scattered photons that are generated during the interaction time in order to further reduce the noise. Furthermore, performing this experiment on a high-power and high-repetition rate laser facility, such as the ELI facility \cite{eli,mourou_extreme_2011} or the Apollon facility \cite{papadopoulos_apollon_2016}, would allow for gathering the statistics essential for verification of the scattering event. The scattering signal would also be boosted by orders of magnitude by performing the experiment on the upcoming \SI{100}{\peta\watt} Station of Extreme Light (SEL) laser facility as may be seen from the scaling in Fig.(3).

Of course, it is practically impossible to achieve a perfect vacuum even in a laboratory setting. Even within a vacuum chamber, there are still some particles that can interact with the lasers and generate some noise. The main interaction contributing to this concern is Compton scattering. However, as shown in \cite{lundstrom_using_2006}, the number of Compton-scattered photons that would possibly be generated in the appropriate frequency band is much smaller than the photon-photon scattering signal we expect. Also, as proven in \cite{jentschura_generation_2011}, the \gls{oam} state of these Compton-scattered photons would either be 0 or the opposite of the state of interest. Thus, using a projective OAM filtering technique \cite{Nicolas_2015}, the already small amount of noise can be reduced by \SI{25}{\decibel} compared to the case without the OAM signature while the signal remains as strong, thus making detection of the signal of interest feasible.

In this letter, we have investigated the effect of orbital  angular momentum  on elastic photon-photon scattering. Using effective field theory, we have derived an expression for the expected generated field. We have shown that the \gls{oam} coupling will provide an additional useful signature for the interaction and will allow for the use of quantum optics techniques to discern them from any background radiation. We have also proposed an experimental setup, which we plan to perform at the SEL laser facility, that would allow for the detection of this effect using high-power lasers. The experimental verification of the polarization of the vacuum will open a new era in strong field, low energy quantum electrodynamics for fundamental physics investigations.

\begin{acknowledgments}
We thank all of the staff of the Central Laser Facility, Rutherford Appleton Laboratory and, in particular, Prof. David Neely for stimulating discussions. This work was supported by UKRI-STFC and
UKRI-EPSRC under Grant Nos. ST/P002048/1 and EP/N509711/1. R.A. acknowledges support from the St Anne's Graduate Development Scholarship scheme.
\end{acknowledgments}

%%%%%%%%%% Merge with supplemental materials %%%%%%%%%%
\pagebreak
%\widetext
\begin{center}
\textbf{\large Supplemental Materials}
\end{center}
%%%%%%%%%% Merge with supplemental materials %%%%%%%%%%
%%%%%%%%%% Prefix a "S" to all equations, figures, tables and reset the counter %%%%%%%%%%
\setcounter{equation}{0}
\setcounter{figure}{0}
\setcounter{table}{0}
\setcounter{page}{1}
%\makeatletter
\renewcommand{\theequation}{S\arabic{equation}}
\renewcommand{\thefigure}{S\arabic{figure}}
\renewcommand{\bibnumfmt}[1]{[S#1]}
\renewcommand{\citenumfont}[1]{S#1}
%%%%%%%%%% Prefix a "S" to all equations, figures, tables and reset the counter %%%%%%%%%%

\subsection{Analysis of the non-linear electromagnetic wave equation}
The Euler-Heisenberg Lagrangian shows that even in the vacuum, there is a non-linear coupling between electromagnetic waves. If one again considers the case with three initial waves, then the electromagnetic fields for the waves can be written in complex notation as 
\begin{equation}
    \mathbf{E}_j(\mathbf{r},t)=E_je^{i(\mathbf{k}_j\cdot \mathbf{r}-\omega_j t)}\mathbf{n}_j(\mathbf{k}_j) 
\end{equation} 
With $n_j(k)$ being the polarization of the field and $j\in\{1,2,3\}$ labeling each individual wave. The electromagnetic wave equation derived from the vacuum Euler-Heisenberg Lagrangian is
\begin{equation}
    \partial^\mu\partial_\mu\mathbf{E} = 4\pi(c^2\mathbf{\nabla}(\mathbf{\nabla}\cdotp\mathbf{P}) - \partial_t(\partial_t\mathbf{P} + c\mathbf{\nabla}\times\mathbf{M}))
\end{equation}
\begin{equation}
    \mathbf{M} = (4\pi)^{-1}\xi(-2(\mathbf{E}^2-\mathbf{B}^2)\mathbf{B} + 7(\mathbf{E}\cdotp\mathbf{B})\mathbf{E}) 
    \label{eq:magnetization}
\end{equation}
\begin{equation}
    \mathbf{P} = (4\pi)^{-1}\xi(2(\mathbf{E}^2-\mathbf{B}^2)\mathbf{E} + 7(\mathbf{E}\cdotp\mathbf{B})\mathbf{B})
    \label{eq.polarization}
\end{equation}
Where $\mathbf{P}$ and $\mathbf{M}$ are the effective vacuum polarization and magnetization respectively, and $\mathbf{E}$ and $\mathbf{B}$ are the total sum of the electric and magnetic fields respectively. It should be noted that since the complex notation is being used for the fields, the squared fields are calculated from their real part (e.g. $\mathbf{E}^2=\Re(E)^2$). From Eq.(\ref{eq:magnetization}) and Eq.(\ref{eq.polarization}), it is clear that the non-linearity is cubic, so the inhomogenous wave equation can rewritten to first order as 
\begin{equation}
\begin{split}
     \partial^\mu\partial_\mu\mathbf{E} = \sum\mathbf{A}_{ijk}E_iE_jE_ke^{i((\mathbf{k}_i+\mathbf{k}_j+\mathbf{k}_k)\cdot\mathbf{r}-(\omega_i+\omega_j+\omega_k)t)}
     \\ + \mathbf{B}_{ijk}E_iE_jE^*_ke^{i((\mathbf{k}_i+\mathbf{k}_j-\mathbf{k}_k)\cdot\mathbf{r}-(\omega_i+\omega_j-\omega_k)t)}
     \\ + \mathbf{C}_{ijk}E_iE^*_jE^*_ke^{i((\mathbf{k}_i-\mathbf{k}_j-\mathbf{k}_k)\cdot\mathbf{r}-(\omega_i+\omega_j-\omega_k)t)}
\end{split}
\label{eq:non-linear}
\end{equation}
It is possible to investigate this equation via harmonic analysis in order to simplify it to a more solvable form. The field of interest, considered in the paper for the convenience of the geometry of its interaction, is the fourth wave that is generated through the photon-photon interaction 
\begin{equation}
    \mathbf{E}_4(\mathbf{r},t)=E_4(\mathbf{r},t)e^{i(\mathbf{k}_4\cdot r-\omega_4 t)}\mathbf{n}_4(\mathbf{k}_4)
\end{equation}
with $\mathbf{k}_4=\mathbf{k}_1+\mathbf{k}_2-\mathbf{k}_3$, $\omega_4=c|\mathbf{k}_4|=\omega_1+\omega_2-\omega_3$ and $E_4(\mathbf{r},t)$ is a slowly varying function of space and time. For this generated field, the left-hand side of Eq.(\ref{eq:non-linear}) can be expanded to first order as
\begin{equation}
\begin{split}
    \partial^\mu\partial_\mu\mathbf{E}_4(\mathbf{r},t)\approx -2i(\omega_4\dfrac{\partial}{\partial t} + c^2\mathbf{k}\cdot\mathbf{\nabla})E_4(\mathbf{r},t) \\ \times e^{i(\mathbf{k}_4\cdot r-\omega_4 t)}\mathbf{n}_4(\mathbf{k}_4)
\end{split}    
\end{equation}
Multiplying both sides of Eq.(\ref{eq:non-linear}) by $e^{i(\mathbf{k}_4\cdot r-\omega_4 t)}$, it is then Fourier transformed in space, thereby leaving only the time derivative of $E(\mathbf{r},t)$. When integrating over time, all the non-resonant terms oscillating at $\omega\neq\omega_4$ vanish and the only relevant non-linear coupling term is the resonant $E_1E_2E_3^*$ term. Hence Eq.(\ref{eq:non-linear}) is simplified to 
\begin{equation}
    \partial^\mu\partial_\mu\mathbf{E}_4 \approx \mathbf{B}_{123}E_1E_2E^*_3e^{i((\mathbf{k}_1+\mathbf{k}_2-\mathbf{k}_3)\cdot\mathbf{r}-(\omega_1+\omega_2-\omega_3)t)}
\end{equation}
which is then solved using the Green's function method as described in the paper.
\subsection{Comparison between photon-photon scattering with and without OAM}
The main advantage to using beams with orbital angular momentum (OAM) in photon-photon scattering is the additional OAM signature that is carried by the generated photons distinguishing them from the background radiation. This is predicted to improve the signal-to-noise ratio of interaction making it much more detectable. To see this, first consider the signal of the generated photons. In the paper, the non-linear equation for an incoming beam OAM was derived by first considering a superposition incident plane waves with an angle $\alpha$ with respect to z-axis and an azimuthal phase shift and then integrating them with respect to the phase. The strength of the coupling with no OAM can be derived by considering case with with $\alpha=0$ and no phase variance. By comparing the predicted number of generated photons through photon-photon scattering for each case (with and without OAM), it can be shown that the number of generated photons is practically unaffected when parameters of the incoming beams (power, duration and focal spot) are held constant. This is intuitive as the effect of OAM would only be to modify the OAM state of the generated photons and not the coupling strength of the photon-photon interaction. 

Of course, high-power lasers typically do not carry OAM and it must be imparted onto them using mode converters such as so-called spiral phase plates (SPP). Even if the coupling strength is not affected, the conversion efficiency is critical as a poorly designed converter would weaken the incoming beam, potentially undermining the positive effect of the additional signature. 

Efficiently generating OAM-carrying beams has been a very active topic of research due to their potential applications in quantum computation and communication, and high-density coding of information \cite{nicolas_quantum_2014,ganic_generation_2002}. Conversion methods adapted to high-power lasers have also been developed \cite{shvedov_efficient_2010,campbell_generation_2012}, with methods such as spiral phase optics having a theoretical efficiency of up to $93\%$ when the waist ratio is optimized \cite{longman_mode_2017}. Spiral phase mirrors are particularly well adapted for high-power lasers as they avoid transmission through dielectric media thereby avoiding any problems with group velocity dispersion (GVD) or B-integral \cite{private_comms}. They can also be made using coatings with high-damage thresholds similar to the ones used on the rest of the optics used to manipulate these high-power beams where dielectric coatings can have fluence damage threshold of over $1 \ \mathrm{J}\cdot\mathrm{cm^{-2}}$ \cite{damage_threshold}. Additionally, high-power beams generally have a large diameter so that their fluence becomes well below the damage threshold for the optical coatings. Tests have already been done using spiral phase mirrors to efficiently generate OAM on 100 TW laser beams \cite{longman_aps}. This means that the conversion of one of the input beams into an OAM-carrying one will not have a detrimental effect on the number of photons generated through photon-photon scattering. 

Considering now the noise, the additional unique OAM signature carried by these photons allows for the use of filtering techniques developed in quantum optics research. In \cite{nicolas_quantum_2015}, a projective filtering method is used to efficiently detect photons carrying a particular OAM signature while other modes are suppressed by around 25 dB. The filtering involves using a highly-efficient blazed fork phase hologram that adds a specific amount of OAM. The goal is to choose the hologram so that the desired OAM mode is converted into a regular Gaussian while all other modes are converted to various other OAM modes. The output is then coupled into a single-mode fiber optic cable itself coupled to a single photon detector. These cables can only transmit Gaussian modes while all other modes decay while propagating. Therefore, at the final output, the specific OAM, corresponding to the photons generated through photon-photon scattering, will be detected while the background noise will be filtered by 25 dB. From all these considerations, the signal-to-noise ratio (SNR) of the interaction using an OAM-carrying initial beam is significantly enhanced compared to that of the regular plane wave.              
\bibliographystyle{plain}

\end{document}